\documentclass[preprint,showpacs,preprintnumbers,amsmath,amssymb,nofootinbib]{revtex4}
\usepackage{bbm}
\usepackage{amsfonts}
\usepackage{booktabs}
\usepackage{mathrsfs}
\usepackage{epsfig}
\usepackage{graphicx}
\usepackage{dcolumn}
\usepackage{bm}
\usepackage{amsmath}
\usepackage{slashed}

\let\jnfont=\rm
\def\NPB#1,{{\jnfont Nucl.\ Phys.\ B }{\bf #1},}
\def\PLB#1,{{\jnfont Phys.\ Lett.\ B }{\bf #1},}
\def\EPJC#1,{{\jnfont Eur.\ Phys.\ Jour.\ C }{\bf #1},}
\def\PRD#1,{{\jnfont Phys.\ Rev.\ D }{\bf #1},}
\def\PRL#1,{{\jnfont Phys.\ Rev.\ Lett.\ }{\bf #1},}
\def\MPLA#1,{{\jnfont Mod.\ Phys.\ Lett.\ A }{\bf #1},}
\def\JPG#1,{{\jnfont J.\ Phys.\ G}{\bf #1},}
\def\CTP#1,{{\jnfont Commun.\ Theor.\ Phys.\ }{\bf #1},}
\def\ZPC#1,{{\jnfont Z.\ Phys.\ C }{\bf #1},}
\def\JHEP#1,{{\jnfont JHEP \ }{\bf #1},}
\def\Rv{\not{\hbox{\kern-1pt $R$}}}
\def\p{\not{\hbox{\kern-3pt $p$}}}

\begin{document}
\preprint{\parbox{1.2in}{\noindent }}

\title{A new approach for detecting compressed bino/wino at the LHC}

\author{Chengcheng Han$^1$, Lei Wu$^2$, Jin Min Yang$^3$, Mengchao Zhang$^3$, Yang Zhang$^3$ \\~ \vspace*{-0.4cm}}
\affiliation{
$^1$ Asia Pacific Center for Theoretical Physics, San 31, Hyoja-dong, Nam-gu,
Pohang 790-784, Republic of Korea \\
$^2$ ARC Centre of Excellence for Particle Physics at the Terascale,
School of Physics, The University of Sydney, NSW 2006, Australia  \\
$^3$ State Key Laboratory of Theoretical Physics, Institute of Theoretical Physics,
Academia Sinica, Beijing 100190, China 
}

\begin{abstract}
In some supersymmetric models like split supersymmetry or models with non-universal
gaugino mass, bino (LSP) and winos (NLSP) may have rather small mass splitting in order to
provide the correct dark matter relic density through bino/wino co-annihilation.
Such a scenario with the compressed bino/wino is difficult to explore at the LHC.
In this work we propose to probe this scenario from
$pp \to j \tilde{\chi}^0_2 \tilde{\chi}^\pm_1$ followed by
$\tilde{\chi}^0_2 \to \gamma \tilde{\chi}^0_1$ and $\tilde{\chi}^\pm_1 \to W^{*}\tilde{\chi}^0_1\to \ell^\pm \nu \tilde{\chi}^0_1$
(this method is also applicable to the compressed bino/higgsino scenario).
Through a detailed Monte Carlo simulation for both the signal and the backgrounds,
we find that for a mass splitting $\Delta M \sim 5-15$ GeV between bino (LSP) and wino (NLSP),
the 14 TeV LHC with luminosity of 500$fb^{-1}$ can probe the wino up to 150 GeV
(the sensitivity can reach  $5\sigma$ for $\Delta M = 5$ GeV and
$2\sigma$ for $\Delta M = 15$ GeV).
We also investigate the dark matter detection sensitivities for this scenario
and find that the planned  XENON-1T(2017) cannot fully cover the parameter space
with wino below 150 GeV allowed by relic density and the LUX limits.
\end{abstract}

\maketitle

\section{introduction}
Supersymmetry (SUSY) is a leading candidate beyond the Standard Model (SM),
which can simultaneously
solve the naturalness problem, explain the cosmic dark matter and achieve the gauge coupling
unification. However, current searches at the LHC have not yet found any evidence of SUSY
particles (sparticles). The first two generation squarks and gluino mass bounds have been
pushed up to TeV region by searching for multi-jets with large missing energy \cite{gluino}.
The light third generation squarks and the non-colored sparticles have also been excluded
in the simplified models \cite{exp-incl-ew,exp-st,exp-sb,stop}.
In addition, the observation of
a 125 GeV Higgs boson \cite{atlas,cms} requires rather heavy stops and/or large stop mixing.
These results seem to indicate some sparticle spectrum (like split-SUSY) not favored by the
naturalness. Therefore, it is imperative to explore all possible corners of SUSY parameter
space and for some special scenarios new search strategies are needed.

In the minimal supersymmetric model (MSSM) with $R$-parity the direct search of sparticles
is mainly influenced by the nature of the lightest sparticle (LSP) and the shape
(compressed or not) of the sparticle spectrum.
In many popular SUSY models the electroweak gauginos are the most likely candidates for the LSP
and the next-to-lightest sparticle (NLSP). The `gold-plated' tri-lepton signature
from the associated production of $\tilde{\chi}^0_2\tilde{\chi}^\pm_1$ is usually expected to
have the best sensitivity for probing the electroweak gaugino sector at the LHC \cite{higgsino-3l}.
The ATLAS and CMS collaborations have performed such a study in the simplified wino-like
chargino/neutralino scenario, whose null results excluded
$m_{\tilde{\chi}^\pm_1,~\tilde{\chi}^0_2} <345$ GeV (ATLAS) \cite{atlas-3l}
and $m_{\tilde{\chi}^\pm_1,~\tilde{\chi}^0_2} <270$ GeV (CMS) \cite{cms-3l}
when the winos decay via intermediate gauge bosons and the LSP is almost a massless bino.
As mentioned above, the tri-lepton sensitivity strongly depends on the mass difference
($\Delta M$) between NLSP and LSP. A large $\Delta M$ is typically required to produce the
hard leptons in the final states of the process
$pp \to \tilde{\chi}^0_2 \tilde{\chi}^\pm_1$ followed by
$\tilde{\chi}^0_2 \to Z^{(*)}(\ell^+\ell^-)\tilde{\chi}^0_1$ and
$\tilde{\chi}^\pm_1 \to W^{(*)}(\ell^\pm \nu)\tilde{\chi}^0_1$.
When $\Delta M$ becomes small, a squeezed spectrum of electroweak gauginos would lead to a
small missing energy and soft leptons. For such a compressed
($\Delta M\sim 1-5$ GeV) electroweak gauginos one may look for
(albeit quite challenging at the LHC)
the mono-jet \cite{monojet,soft-lepton}, mono-photon \cite{tata} or mono-$Z$ \cite{carpenter,liutao}
final states, where a visible jet, photon or $Z$-boson
form initial state radiation (ISR) is employed to trigger the process.
In addition, an alternative searching strategy
$pp \to \tilde{\chi}^0_2 \tilde{\chi}^0_3 \to \ell^+ \ell^- + \gamma + \slashed E_T$ has also been
proposed to detect the compressed bino-higgsino in \cite{bino-higgsino}, where the dilepton comes
from the three body decay $\tilde{\chi}^0_{2,3} \to \ell^+ \ell^- \tilde{\chi}^0_1$ and the photon
is from the loop-induced decay $\tilde{\chi}^0_{2,3} \to \gamma \tilde{\chi}^0_1 $.
But this method is suitable only for a `well-forged' mass difference $\Delta M \sim 25-70$ GeV.

In this work we propose a new search channel $pp \to j\tilde{\chi}^0_2 \tilde{\chi}^\pm_1$
followed by $\tilde{\chi}^0_2 \to \gamma \tilde{\chi}^0_1$ and
$\tilde{\chi}^\pm_1 \to W^{*}(\ell^\pm \nu)\tilde{\chi}^0_1$ to probe the compressed
($\Delta M \sim 5-20$ GeV) bino (LSP)-winos (NLSP) at the LHC.
Note that such a compressed scenario may readily happen in
split-SUSY or models with non-universal gaugino masses at the boundary where the correct dark matter
relic abundance can be achieved from the bino-wino co-annihilation \cite{well-temper,tata2}.
The final states of this channel are characterized
by a hard jet with a soft photon and lepton: $j+\gamma+\ell^\pm+\slashed E_T$.
The detectability of this channel comes from three sides:
(1) For the cross section of $j\tilde{\chi}^0_2 \tilde{\chi}^\pm_1$, the winos usually has larger cross section than other electroweakinos;
(2) The chargino $\tilde{\chi}^\pm_1$ dominantly decays (with a branching ratio near 100\%) into the
neutralino LSP plus a virtual $W$-boson (comparatively the decays
$\tilde{\chi}^0_{2,3} \to \ell^+ \ell^- \tilde{\chi}^0_1$ have very small branching ratios);
(3) A hard ISR jet in the final states can avoid the conventional huge electroweak
$W\gamma$ background that is the most severe handicap for the
$\tilde{\chi}^0_2 (\to \gamma \tilde{\chi}^0_1)\tilde{\chi}^\pm_1(\to \ell^\pm \nu \tilde{\chi}^0_1)$
production. On the contrary, the background $Wj\gamma$ can be efficiently removed by requiring large missing energy.

The rest of this work is organized as follows.
In Sec.2 we study the observability of this channel at the 14 TeV LHC.
In Sec.3 we show the dark matter direct detection sensitivity to the compressed
bino-wino scenario. Finally, we draw our conclusion in Sec.4.

\section{Probing compressed bino/wino at LHC}
\subsection{The production channel and its signal}
\begin{figure}[htbp]
\includegraphics[width=5cm,height=3.0cm]{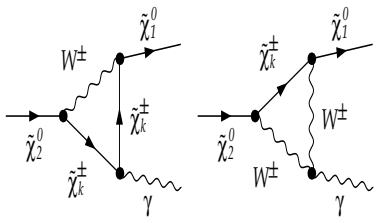}
\vspace*{-.6cm}
\caption{The relevant Feynman diagrams for the loop decay $\tilde{\chi}_2^0\rightarrow \gamma \tilde{\chi}_0^1$ in our study. Here $k=1,2$.}
\label{fig1}
\end{figure}

\begin{figure}[ht]
\includegraphics[width=10cm]{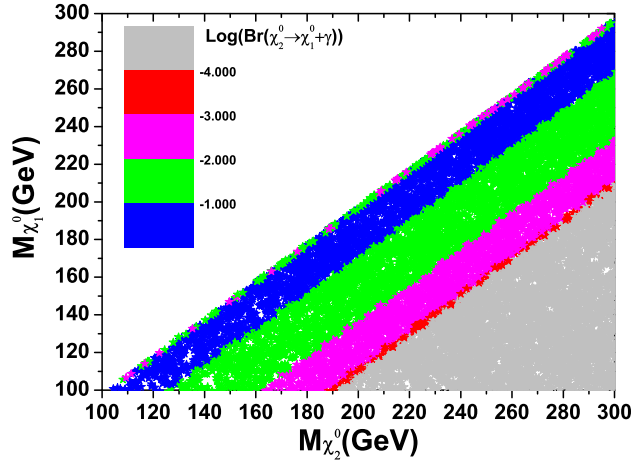}
\vspace*{-0.5cm}
\caption{The dependence of branching ratio of $\tilde{\chi}_2^0\rightarrow \gamma \tilde{\chi}_0^1$ on the masses of $\tilde{\chi}_{1,2}^0$. We use \textbf{SUSY-HIT} \cite{susyhit} to calculate the branching ratio. The gluino and higgsinio mass parameters $M_3$ and $\mu$ are assumed at 2 TeV and 1 TeV, respectively. We vary parameter $\tan\beta$ from 3 to 50 and the gaugino masses $M_{1,2}$ from 100 GeV to 300 GeV. All the samples are required to satisfy the Higg mass $125\pm2$ GeV. The slepton and the first two generation squark sectors are set a common mass $M_{SUSY}=2$ TeV.}
\label{fig2}
\end{figure}

Concentrating on a small mass splitting between bino and winos ($\lesssim$ 20 GeV), we consider
the $j+\ell+\gamma+ \slashed E^{miss}_T$ signal from the $\tilde{\chi}^\pm_1\tilde{\chi}^0_2j$ production
followed by the decays
\begin{eqnarray}
&& \tilde{\chi}^\pm\rightarrow W^*\tilde{\chi}^0_1\rightarrow \ell \nu \tilde{\chi}^0_1 , \nonumber\\
&&  \tilde{\chi}_2^0\rightarrow \gamma \tilde{\chi}_0^1
\label{produciton}
\end{eqnarray}
where the decay channel $\tilde{\chi}_2^0\rightarrow \gamma \tilde{\chi}_0^1$ is a loop process \cite{n2-decay}, as shown in Fig.\ref{fig1}.
Note that the branching ratio of this decay is sensitive to the mass splitting of $\tilde{\chi}_2^0$ and $\tilde{\chi}_0^1$.
As shown in Fig.\ref{fig2}, when the mass splitting is small enough, this decay has a sizable branching ratio because the tree level three-body decay
$\tilde{\chi}_2^0\rightarrow Z^* \tilde{\chi}_0^1\rightarrow f \bar{f} \tilde{\chi}_0^1$ is suppressed.
From Fig.\ref{fig2} we see that for $\Delta m < 20$ GeV, the decay $\tilde{\chi}_2^0\rightarrow \gamma \tilde{\chi}_0^1$
can have a branching ratio as large as $10\%$. As the mass splitting gets large, the decay branching ratio
reduces rapidly. We should also note that when the mass splitting is very small ($< 5$ GeV),
the decay branching ratio will reduce a bit. This is because for such a tiny mass splitting $\tilde{\chi}_2^0$
would have sizable bino component which will suppress the coupling of $\tilde{\chi}_2^0 \tilde{\chi}_1^+ W$ in the loops
of Fig.\ref{fig1}.

In the following we choose four benchmark points to perform Monte Carlo simulation.
These points are listed in Table \ref{tab1}.
Here the masses of $\tilde{\chi}^0_{1,2}$ are obtained by tuning the bino and wino masses.

\begin{table}[th] \centering\caption{Four benchmark points used in our Monte Carlo simulation.
Here $\mu$ and $\tan\beta$ are set to be 1 TeV and 30, respectively.}
\begin{tabular}{|c|c|c|c|c|c|c|c|}
\hline  ~~($m_{\tilde{\chi}^0_1}$, $m_{\tilde{\chi}^0_2}$) in GeV~~      &    ~~(130,150)~~ & ~~(135,150)~~ & ~~(140,150)~~ & ~~(145,150)~~  \\
\hline  ~~Br($\tilde{\chi}^0_2\rightarrow \tilde{\chi}^0_1 \gamma$)~~      &0.101 	&0.2266 	&0.495 	&0.834 	  \\
\hline
\end{tabular}
\label{tab1}
\end{table}

\subsection{Monte Carlo simulation for the signal and backgrounds}
Our signal is a soft lepton, a soft photon, a hard jet plus large missing energy.
The largest background is from the $W\gamma j$ production (the $W$-boson here can be on-shell or virtual).
To suppress this background, we require a rather hard jet $P_{T} > 300$ GeV and a large missing energy.
Although this requirement can also hurt our signal, the S/B ratio can be enhanced.
In addition, since the photon and lepton are typically soft in our signal,
we require them to be softer than 40 GeV and  25 GeV, respectively.
Our selection criteria are summarized as
\begin{itemize}

\item[(i)] One hard jet.  We require a hard jet with $P_{T}(j_1) >300$ GeV and $|\eta({j_1})| <$ 2.5 (not $b$-tagged).
Any events with more visible jets ($P_T(j_2) > 30$ GeV, $|\eta(j_2)| <$ 2.5) will be vetoed.

\item[(ii)] Large $\slashed E^{miss}_T$. The final state has two massive LSP neutralinos, which recoil to the leading jet and then induce
a large missing energy. Here we require $\slashed E^{miss}_T > 300$ GeV.

\item[(iii)] One isolated lepton. We require an isolated electron with $P_T^e > 10$ GeV,
$|\eta^e| <$  1.37 (or 1.52 $< |\eta^e| <$ 2.47) or an isolated muon with $P_T^\mu >10$ GeV,
$|\eta^\mu|  <$ 2.4. An upper limit cut of  25 GeV is also imposed on the lepton $P_T^{e,\mu}$.

\item[(iv)] One isolated photon. We require an isolated photon with $|\eta^\gamma| <$  2.37 (excluding
the calorimeter transition region 1.37 $< |\eta^\gamma| <$ 1.52) and $P_T^\gamma >$ 10 GeV.
Also an upper limit cut of 40 GeV is imposed on the photon $P_T^\gamma$.
\end{itemize}

\begin{figure}[htbp]
\includegraphics[width=12cm]{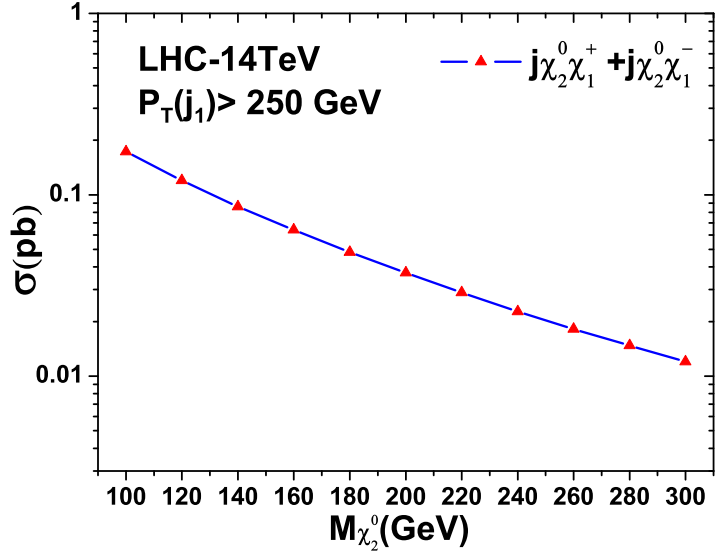}
\vspace*{-0.5cm}
\caption{The dependence of the cross section of $\tilde{\chi}^\pm_1\tilde{\chi}^0_2j$ on the $m_{\tilde{\chi}^0_2}$ at 14 TeV LHC.}
\label{fig3}
\end{figure}
Fig.\ref{fig3} shows the production rate of $\tilde{\chi}^\pm_1\tilde{\chi}^0_2j$ at the 14 TeV LHC for different wino masses. Here the cross section is calculated at tree level by using \textsf{MadGraph5} (in order to obtain reasonable statistics, we imposed a cut $p_T({j_1}) >$ 250 GeV on the first leading jet for signals and  backgrounds
in the parton-level events generation). From this figure we see that the production rate can reach 0.1 pb when the $\tilde{\chi}^0_2$ mass is below 130 GeV.
In our benchmark points where the $\tilde{\chi}^0_2$ mass is 150 GeV, the cross section can reach 0.074 pb before multiplying the decay branching ratios.

As mentioned above, the dominate background is from the $W\gamma j$ production whose cross section is about 0.75 pb at tree level (with $P_T(j_1)>$ 250 GeV).
Another sizable background is $Z(\tau\tau)$+jets with $\tau$ decays in the lepton channel and one of the leptons mistagged as a photon (the large missing energy can arise from the neutrinos from the $\tau$ leptonic decays). Although the $\tau$ decay branching ratio and the efficiency of a lepton misidentified as a photon can suppress the $Z(\tau\tau)$+jets background, this background cannot be ignored because its production rate is quite large (0.84 pb with $P_T(j_1)>$ 250 GeV).

About other possible backgrounds we have the following comments:
\begin{itemize}
\item[$\bullet$] W+jets background. The ATLAS performed the measurement of $W\gamma$ production \cite{wa}
where W+jets is the background. In the experiment an inclusive measurement is performed after requiring:
exactly one lepton with $P_T > 25$ GeV, at least one isolated photon with $E_T^{\gamma} >15$ GeV
and $\slashed E_T^{miss}$ above 35 GeV. Other selection requirements include: $m_T$ should be larger than 40 GeV
and $m_{e\gamma}$ is not within 15 GeV of $Z$-boson mass.
We note that the W+jets events can mimic $W\gamma$ signal when a jet is mistagged as a photon.
We calculate the $\sigma_{W+jets}/\sigma_{W\gamma}$ value after these cuts (for W+jets events we
choose a jet as a photon) and compare with the final counting events in the experiment.
We find the efficiency of the jets mistagged as a photon to be $\sim10^{-4}$.
We recalculate the $\sigma_{W+jets}/\sigma_{W\gamma j}$ value at 14 TeV after our cuts and
find it $\lesssim 100$. After multiplying the mistagged factor we find that the W+jets background
is much smaller than the $W\gamma j$ background.
\item[$\bullet$] Top quark background. The $t\bar{t}$ background can mimic our signal when top quarks
decay leptonically and one of the leptons is mis-tagged as a photon.
We calculate the $t\bar{t}$ cross section under the requirement $\slashed E_T^{miss} >$200 GeV at parton level
and find it to be 0.32 pb. The cut efficiency with $\slashed E_T^{miss} >$ 300 GeV and $P_T(j_1) >$ 300 GeV (for the
leading jet) is about 0.1. The probability to find a pre-selection lepton and photon is 0.02.
The cut efficiency for the lepton and photon is 0.15 and 0.3, respectively.
The jet veto and $b$-jet veto takes factor more than 40.
After all the cuts, this background is approximately smaller than 0.001 fb.
So we can safely ignore this background.
Another background is $t\bar{t}\gamma$ whose production rate is 4 fb after requiring $\slashed E_T^{miss}>$ 200 GeV
at parton level. The efficiency with all the cuts is $6.0\times 10^{-4}$. So its contribution is also
negligible.
\item[$\bullet$] WW/ZZ/ZW. The production rate of these gauge bosons is at fb level
after requiring a leading jet harder than 250 GeV and a factor smaller than 0.06 should be multiplied
to include the lepton mistagging as a photon.
\item[$\bullet$] $Z\gamma j$. This background has a fb production rate after requiring a leading
jet larger than 250 GeV and can be suppressed by requiring large $E_T^{miss}$.
So it can be safely neglected.
\end{itemize}
Other backgrounds like Z+jets when Z decays to electrons or muons and a lepton can be faked as a photon and the $\gamma$+jets background when a jet is misidentified as a lepton, are highly suppressed by the requirement of $\slashed E_T^{miss}>$ 300 GeV and also by a small mistag factor.

We use \textsf{MadGraph5} \cite{mad5} to simulate the signal and the backgrounds. We carry out the parton shower and the fast detector simulation through \textsf{PYTHIA} \cite{pythia} and \textsf{Delphes} \cite{delphes}, respectively. We also use the anti-$k_t$ algorithm \cite{anti-kt} to cluster jets and match our matrix element with parton shower in MLM scheme \cite{mlm}. The cross sections of the signal and backgrounds are evaluated at tree level in our simulation.

In Table \ref{tab2} we show the cut flows of backgrounds and signal. We see that the background $W\gamma j$ can be suppressed by requiring the lepton and photon to be soft.
With all the cuts the efficiency of the signal is about an order larger than the $W\gamma j$ background.

\begin{table}[th] \centering\caption{The cut efficiency for the backgrounds and the signal
$j+\ell+\gamma+ \slashed E^{miss}_T$. The signal efficiency is displayed for the four benchmark points listed in
Table \ref{tab1}.}
\begin{tabular}{|c|c|c|c|c|c|c|c|}
\hline  ~~cuts ~~             & W$\gamma$j  & Z($\tau\tau$)+j       &(130,150) & (135,150)& (140,150) & (145,150) \\
\hline  an isolated lepton $p_T^{\ell}>$ 10 GeV    	        &51.9\% 	&28.5\% 	&35.7\%    &31.7\% 	&25.9\% 	 &16.1\% 	   \\
\hline    $p_T^{\ell}<$25 GeV    	            &5.5\% 	    &5.74\% 	&20.4\%    &20.89\% 	&20.0\% 	&14.3\% 	   \\
\hline  an isolated photon $p_T^\gamma>$10 GeV              &3.4\% 	    &1.56\% 	&14.8\%    &14.3\% 	&11.7\% 	 &6.2\% 	   \\
\hline  $p_T^\gamma <$40 GeV    	            &1.1\% 	    &0.3\% 	&7.9\%     &9.0\% 	&8.4\% 	     &4.8\% 	   \\
\hline
$
\begin{array}{ll}
 P_T(j_1) > 300 \textrm{GeV }  \\
\textrm{  (veto additional jets)}
\end{array}
$
&0.26\%  &0.044\% 	&1.9\%     &2.2\% 	&2.1\% 	     &1.32\% 	  \\
\hline   $\slashed E_T^{miss} >300$ GeV                       &0.15\% 	&0.004\% 	&1.5\%     &1.8\% 	&1.87\% 	 &1.28\%    \\
\hline
\end{tabular}
\label{tab2}
\end{table}

\subsection{Statistical significance at the LHC}
\begin{table}[th] \centering\caption{The cross sections of the signal and backgrounds at the 14 TeV LHC
with all the cuts. The statistical significance with 300 fb$^{-1}$ and 500 fb$^{-1}$ are also shown.
The signal is displayed for the four benchmark points listed in Table \ref{tab1}.}
\begin{tabular}{|c|c|c|c|c|c|c|}
\hline   ~ & W$\gamma$j (fb)  &Z($\tau\tau$)+j (fb) & Signal (fb)& S/B & S/$\sqrt{B}$ ($300fb^{-1}$)  &  S/$\sqrt{B}$ ($500fb^{-1}$)\\
\hline  (130,150)    	&1.14 	&0.03 	 	&0.04 	&0.03 	&0.58  & 0.75 \\
\hline  (135,150)      &1.14 	&0.03 	 	&0.10 	&0.09 	&1.66  & 2.15 \\
\hline  (140,150)      &1.14 	&0.03 	 	&0.22 	&0.19 	&3.54  &  4.57\\
\hline  (145,150)      &1.14 	&0.03 	 	&0.26 	&0.22 	&4.16  &  5.38\\
\hline
\end{tabular}
\label{tab3}
\end{table}
Finally, in Table \ref{tab3} we show the cross sections and the statistical significance after all cuts.
From this table we see that with all the cuts the signal with a small mass splitting
has a good S/B and could be probed at the future LHC. Since the cut efficiency of the four benchmark points does not vary greatly,
the results largely depend on the production rate of the signal.
In other words, the decay branching ratio $\tilde{\chi}_2^0\rightarrow \gamma\tilde{\chi}_1^0$ determines the observability.
It is clearly shown that the benchmark point with the smallest mass splitting gives the best result,
whose statistical significance can reach $4\sigma$ for 300 $fb^{-1}$ and 5$\sigma$ for 500 $fb^{-1}$.
When the mass splitting is enlarged to 15 GeV, the sensitivity can still reach $2\sigma$
for 500 $fb^{-1}$.

\section{Probing compressed bino/wino in dark matter detection}
As a complementary search for the electroweak gauginos, we also investigate the
sensitivity of dark matter direct detection experiments to the compressed bino/wino.
Here we focus on the bino LSP and wino NLSP case.
In order to account for the dark matter relic density, we
scan the parameter space in the following ranges:
\begin{eqnarray}
100 ~\rm{GeV} <M_1< 300 ~\rm{GeV}, ~~100 ~\rm{GeV}<M_2< 300 ~\rm{GeV}, ~~3 < \tan\beta < 50
\end{eqnarray}
Other SUSY mass parameters except the stop sector are fix at 2 TeV ($\mu$ is fixed at 1 TeV).
The parameters in the stop sector are scanned in the following ranges
\begin{eqnarray}
700 ~\textrm{GeV} <(M_{\tilde{Q}_3},M_{\tilde{t}_R})< 2 ~\textrm{TeV},
~~-3 ~\textrm{TeV}< A_t < 3~\textrm{TeV}
\end{eqnarray}
where we choose the lower bound by considering the direct stop search limit. In our scan we consider the following constraints:
\begin{itemize}
\item[(1)]  The SM-like Higgs mass in the range of 123-127 GeV.
We use \textsf{FeynHiggs2.8.9} \cite{feynhiggs} to calculate the Higgs mass and impose the experimental
constraints from LEP, Tevatron and LHC by \textsf{HiggsBounds-3.8.0} \cite{higgsbounds}.
\item[(2)]  Various $B$-physics constraints at 2$\sigma$ level.
We implement the constraints by using the package \textsf{SuperIso v3.3} \cite{superiso}, including
$B\rightarrow X_s\gamma$ and the latest measurements of $B_s\rightarrow \mu^+\mu^-$,
$B_d\rightarrow X_s\mu^+\mu^-$ and $ B^+\rightarrow \tau^+\nu$.
\item[(3)] The constraints from the precision electroweak observables such as
$\rho_\ell$, $\sin^2 \theta_{\rm eff}^\ell$, $m_W$ and $R_b$ \cite{rb} at $2\sigma$ level.
\item[(4)] The thermal relic density of the lightest neutralino (as the dark matter candidate)
is within the 2$\sigma$ range of the Planck value \cite{planck}.
We also consider the direct dark matter search limits from LUX \cite{LUX}.
We use the code \textsf{MicrOmega v2.4} \cite{micromega} to calculate the relic abundance
and DM-nucleon scattering.
\end{itemize}

\begin{figure}[htbp]
\includegraphics[width=18cm]{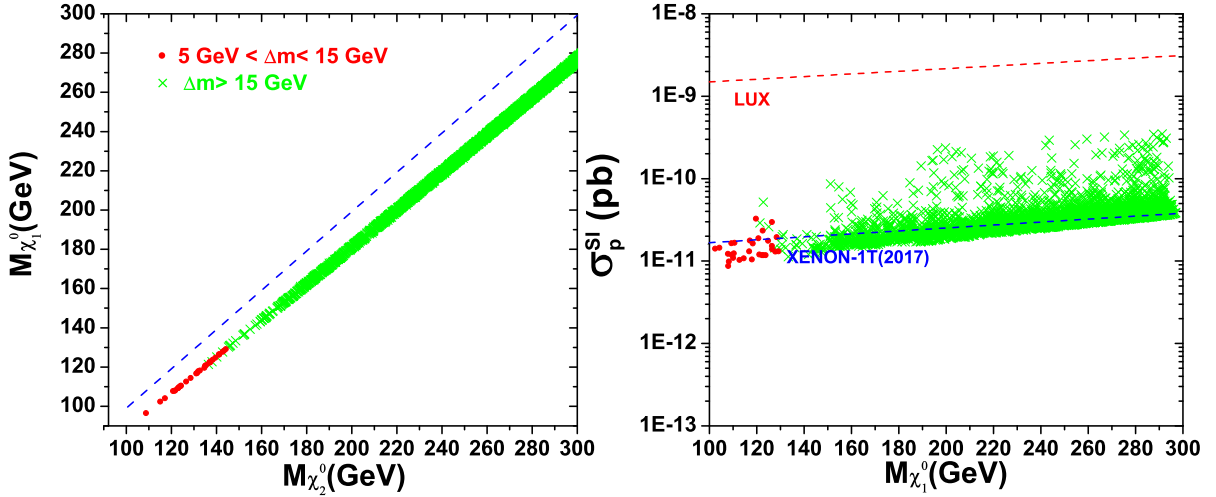}
\vspace*{-1cm}
\caption{The scatter plots of the samples allowed by the $2\sigma$ dark matter relic density
and various collider constraints listed in the context. The left panel shows the plots in
the plane of $m_{\tilde{\chi}_1^0}$ versus $m_{\tilde{\chi}_2^0}$ while the right panel shows the neutralino
dark matter-nucleon scattering cross section versus the neutralino mass.}
\label{fig4}
\end{figure}
In Fig.\ref{fig4} we show the scatter plots of the samples allowed by the $2\sigma$ dark matter relic density
and various collider constraints (1)-(4) listed above. From Fig.\ref{fig4}, it can be seen that, for our region with $\tilde{\chi}_2^0 \lesssim 150$ GeV, the dark matter relic density can be guaranteed by the co-annihilation among $\tilde{\chi}_2^0$, $\tilde{\chi}_1^\pm$ and $\tilde{\chi}_1^0$ due to their small mass splitting $\Delta M \sim 5-15$ GeV. However, such a region could not be covered by the current LUX and future XENON-1T(2017) experiments because of the suppression of the coupling. So, our proposed method can be served as a complementary way to probe this compressed bino/wino scenario at the LHC.

Finally, we would like to stress that in our analysis we worked in the framework of MSSM, in which
the neutralino LSP (dark matter candidate) is bino-like ($M_1<M_2<\mu$). In some extensions of the
MSSM, for example, the popular next-to-minimal supersymmetric model, the property of the neutralino LSP
(dark matter candidate) may be quite different because it tends to be singlino-like
and rather light \cite{cao-nmssm}.
Then the bino-wino co-annihilation is no longer the mechanism to give the correct dark matter relic
abundance and also the signature of the production $pp \to j \tilde{\chi}^0_2 \tilde{\chi}^\pm_1$ is different from
what we studied. Also, some recent studies \cite{citenew} discussed the phenomenology of
electroweak gauginos in other miscellaneous scenarios.

\section{Conclusion}
In this work we proposed to use the signal
$\ell+j+\gamma+\slashed E_T^{miss}$ to probe the compressed bino/wino scenario at the LHC.
From detailed Monte Carlo simulations we find that the 14 TeV LHC with luminosity of 500 fb$^{-1}$
can probe the wino NLSP up to 150 GeV for a wino-bino mass splitting 5-15 GeV.
Such a method is also applicable to the compressed bino/higgsino scenario. We investigated
the dark matter detection sensitivities for this scenario
and found that the planned  XENON-1T(2017) cannot fully cover the parameter space
with wino below 150 GeV allowed by relic density and the LUX limits.

\section*{Acknowledgement}
This work was supported by the Korea Ministry of Education, Science and Technology (MEST)
under the Young Scientist Training Program at the Asia Pacific Center for Theoretical Physics (APCTP),
by the Australian Research Council, by the National Natural Science Foundation of China (NNSFC) under
grants Nos. 11222548, 11305049, 11405047, 11275245, 10821504, 11135003
and by the Specialized Research Fund for the Doctoral Program
of Higher Education under Grant No.20134104120002.

\end{document}